\RequirePackage{lineno} 
\documentclass{revtex4}

\usepackage{graphicx}
\usepackage{dcolumn}
\usepackage{bm}
\usepackage{color}
\usepackage{eso-pic}
\usepackage{amsmath}

\newcommand{\YBCO}{$\mbox{YBa}_2$$\mbox{Cu}_3$$\mbox{O}_7${$\mbox{ } $}}
\begin{document}

\title{Toward terahertz heterodyne detection with superconducting Josephson nano-junctions}

\author{A. Luo$^1$, T. Wolf$^1$, Y. Wang$^1$, M. Malnou$^1$,  C. Ulysse$^2$, G. Faini$^2$, P. Febvre$^3$, M. Sirena$^4$, J. Lesueur$^1$, N. Bergeal$^1$}

\affiliation{$^1$Laboratoire de Physique et d'Etude des Mat\'eriaux - UMR8213-CNRS, ESPCI ParisTech, 10 Rue Vauquelin - 75005 Paris, France.}
\affiliation{$^2$Laboratoire de Photonique et de Nanostructures LPN-CNRS, Route de Nozay, 91460 Marcoussis, France.}
\affiliation{$^3$IMEP-LAHC - UMR 5130 CNRS, UniversitŽ de Savoie, 73376 Le Bourget du Lac cedex, France.}
\affiliation{$^4$Centro At\'omico Bariloche, Instituto Balseiro Ð CNEA and Univ. Nac. de Cuyo, Av. Bustillo 9500, 8400 Bariloche, Rio Negro Ð Argentina.}
\date{\today}

\begin{abstract}
  In this letter, we present the study of the high-frequency mixing properties of ion irradiated \YBCO Josephson nano-junctions. The frequency range, spanning above and below the characteristic frequencies $f_c$ of the junctions, permits a clear observation of the transition between two mixing regimes. The experimental conversion gain was found to be in good agreement with the prediction of the  three ports model. Finally, we discuss the potential of the junctions to build a Josephson mixer operating in the terahertz frequency range.
\end{abstract}

\maketitle
 
The terahertz (THz) region of the electromagnetic spectrum [0.3-10THz] has, so far, not been exploited fully due to the lack of suitable sources and detectors \cite{tonouchi}. Indeed, THz frequency lies between the frequency range of traditional electronics and photonics where the existing technology cannot be simply extended. Low temperature superconducting-insulating-superconducting Niobium tunnel junctions are currently used as frequency mixing element in heterodyne receivers \cite{graauw}, providing extremely low noise and high sensitivity \cite{mears}. However, they are intrinsically limited in frequency by the energy gap of Nb ($\sim$800 GHz ) and operate only at low temperature (4.2K). \\
\indent An alternative to these devices consists of using high-temperature superconducting (HTS)  receivers based on Josephson Junctions mixers or hot-electron bolometers. In addition to the obvious advantage of a much higher operating temperature, the investigation of  HTS mixers is motivated by the possibility of approaching  quantum-limited noise performance at frequencies higher than what is possible with conventional Nb devices \cite{schoelkopf,classen}. Hence, it is important to develop HTS devices and related heterodyne mixer technology for applications in the THz range. However, despite a few promising realizations mainly based on grain-boundary or ramp edge junctions \cite{scherbel,harnack,tarasov}, the development was slowed down by the difficulty to build a junction technology sufficiently reliable to fabricate complex devices. Over the past years, a new approach based on ion irradiation has been developed to make Josephson nano-junctions with high temperature superconductors. This method has been used to make reproducible junctions \cite{bergeal,kahlmann}, SQUIDs \cite{bergealsq} and  large scale integrated Josephson circuits \cite{cybart1, cybart}. In this Letter, we present a study of the high-frequency mixing properties of Josephson nano-junctions made by this technique. \\

\indent Fig 1a shows an optical picture of a Josephson junction embedded in a self-complementary THz spiral antenna. We used 70 nm thick \YBCO films grown on sapphire substrates. The junction is defined in a 2 $\mu$m wide superconducting channel by irradiating through a 20 nm wide slit with 100 keV oxygen ions at a fluence of 3$ \times$10$^{13}$ at/cm$^2$. The fabrication method has been described in detail in reference \cite{bergealjap}. The junction is connected to contact pads for dc biasing and to a microwave transmission line to read the output signal. The back side of the sapphire substrate is placed in contact with a silicon hyper-hemispheric lens located at the focal point of a parabolic mirror exposed to external signals though the window of the cryostat. The output signal at intermediate frequency is amplified by a cryogenic HEMT amplifier [4-8GHz] before further amplification at room temperature. An isolator is placed in the chain to  minimize the back-action of the amplifier on the Josephson mixer.\\
\indent The  resistance of the junction as a function of temperature is shown in figure 1b. The highest transition refers to the superconducting transition of the reservoirs (i.e. non irradiated electrodes ) at $T^0_{c}$  = 85 K, which corresponds to the transition temperature of the unprocessed \YBCO film  \cite{bergeal}. The second transition at the lower temperature T$_{J}$ = 68 K corresponds to the occurrence of the Josephson coupling between the two electrodes, and not to the transition of the irradiated part itself which is expected at the lower temperature $T'_c\approx$ 52 K.   Junctions fabricated by this method are in the low capacitance regime, defined by a McCumber dimensionless parameter $\beta_c=\frac{2e}{\hbar}I_cR_n^2C$ much smaller than one \cite{mccumber} ($I_c$ is the critical current, $R_n$ the normal-state resistance and $C$ the capacitance of the
junction). Junctions have non-hysteretic current-voltage characteristics with an upward curvature of the dissipation branch and no sharp feature at the gap voltage (fig1b inset), a behavior well described by the Resistively Shunted Junction (RSJ) model \cite{mccumber}. \\
\indent  Below T$_{J}$, the critical current $I_c$ increases with a quadratic law when the temperature is lowered (fig1b).  A the lower temperature $T'_c$,  the junction enters a flux-flow regime and the I(V) characteristics display a downward curvature (inset fig1b). The characteristic frequency $f_c$ of the junction is defined by the $I_cR_n$ product via the Josephson frequency $f_c = (2e/h)I_cR_n$.  Although the mixing operation is optimal when the signal frequency is lower than $f_c$,  it can be performed up to frequencies corresponding to several times the value of $f_c$ with a reduced conversion efficiency. As can be seen in fig 1b, the characteristic frequency of the nano-junction is temperature-dependent and takes a maximum value $f_c^{\mathrm{max}}$= 75GHz at 58K. The decrease of $f_c$ below this temperature is due to the drop of $R_n$ as it approaches the flux flow regime.\\
The observation of Shapiro steps is important to evaluate the dynamic properties of a Josephson junction \cite{shapiro}. I(V) characteristics under 20GHz radiation have been measured as a function of radiation power. Figure 2a shows the differential resistance of the junction $\frac{dV}{dI}$ as a function of bias current and power radiation. The oscillations of several Shapiro steps with power can be clearly observed here. The data are in qualitative agreement with the predictions of the current driven RSJ model which admits a single parameter $f_c$ (figures 2b and 2c).  \\

 To study the mixing properties of the junction, a strong local oscillator signal at frequency $f_{\mathrm{LO}}$ and a weaker test signal at frequency $f_{\mathrm{s}}$ are injected through the optical window of the cryostat.  Three different ranges of frequency have been investigated : (i)  $f_{\mathrm{LO}}$=20GHz $<f_c^{\mathrm{max}}$, (ii)  $f_{\mathrm{LO}}$=70GHz $\approx f_c^{\mathrm{max}}$ and (iii) $f_{\mathrm{LO}}$=140GHz $>f_c^{\mathrm{max}}$. Figure 3 shows the output power measured at the intermediate frequency (IF) $f_{\mathrm{IF}}=|f_{\mathrm{LO}}-f_{s}|$=5.5GHz as a function of the dc voltage $V$ across the junction. In these measurements, the power of the local oscillator has been set to reduce the critical current to approximatively half of its value as it corresponds to an optimal operation point for mixer performances. In the regime $f_{\mathrm{LO}}<f_c^{\mathrm{max}}$, the power at $f_{\mathrm{IF}}$ displays maximums located at the center of each Shapiro step whereas in the regime $f_{\mathrm{LO}}>f_c^{\mathrm{max}}$ there are two maximums within a Shapiro step, separated by a dip at the center. The regime $f_{\mathrm{LO}}\approx f_c^{\mathrm{max}}$ corresponds to an intermediate situation where the power at $f_{\mathrm{IF}}$ is approximatively flat at the center of the steps. Mixing at frequency higher than 140 GHz was not investigated in this study.\\

  The three ports model has been used in the context of the RSJ model to calculate the theoretical performance of the mixer \cite{taur,schoelkopf}. It describes the linear response due to a small signal by solving the non-linear response of the mixer to the local oscillator illumination. Only three frequencies of relevance are considered: the upper side-band (USB) $f_{\mathrm{LO}}+f_{\mathrm{IF}}$, the lower side-band (LSB) $f_{\mathrm{LO}}-f_{\mathrm{IF}}$ and  $f_{\mathrm{IF}}$. The frequency conversion matrix is defined by
\begin{displaymath}
 \left( \begin{array}{ccc}
V_u\\
V_0\\
V_l^*\\
\end{array} \right)=
\left( \begin{array}{ccc}
Z_{uu}& Z_{u0} & Z_{ul} \\
Z_{0u} & Z_{00} &  Z_{0l}  \\
 Z_{lu} & Z_{l0}& Z_{ll}\\
\end{array} \right)
 \left( \begin{array}{ccc}
I_u\\
I_0\\
I_l^*\\
\end{array} \right)
\end{displaymath}
   
where the $u$, $l$, $0$ stand for USB, LSB and IF respectively, and $V_j$ and $I_i$ are the voltages and currents at those respective frequencies. Each element $Z_{ij}$ is simply the ratio of the voltage at the relevant frequency $V_j$ with the current injected $I_j$. The diagonal elements are the impedances of the junction at the corresponding frequencies whereas  the off-diagonal elements give the non-linearity necessary for mixing. The matrix elements can be calculated by including small signals at the USB and IF in the RSJ model pumped by the LO signal.  The thermal noise is included in the model by the addition of an uncorrelated Gaussian distributed random current fluctuation of variance $\sigma^2=\Gamma/\Delta\tau$, where $\Gamma=2ek_BT/\hbar I_c$ is the dimensionless RSJ noise parameter and $\Delta\tau$ is the normalised time step. The matrix is calculated for each value of the bias current  and averaged over many realisations of the matrix to take into account the thermal noise. In order to determine the conversion efficiency, we introduce the external part of the circuit described by  the diagonal impedance matrix $Z_{\mathrm{ext}}$ whose  elements $Z_u$, $Z_l$ and $Z_0$ are connected to the mixer inputs. Here $Z_u$ and $Z_l$ represent the impedance of the spiral antenna (80$\Omega$) and are taken to be identical. $Z_0$ is the 50$\Omega$ impedance of the microwave readout line. The conversion efficiency is defined as the ratio of the IF power dissipated in the impedance $Z_0$ to the available test signal power on $Z_u$ (or $Z_l$). It can be derived from the conversion matrix\cite{taur}
   \begin{eqnarray}
   \eta_c=4*\Re(Z_u)\Re(Z_0) |Y_{0u}|^2
   \label{conversion}
   \end{eqnarray}
where $Y_{0u}$ is the matrix element of the $Y$ matrix defined by $Y=(Z+Z_{\mathrm{ext}})^{-1}$\\
\indent As shown in figure 3d, the experimental data are in good agreement with the theoretical calculations derived from the three ports model \cite{taur}.  In particular, it describes well the crossover from the first regime $f_{\mathrm{LO}}<f_c^{\mathrm{max}}$ to the second regime $f_{\mathrm{LO}}>f_c^{\mathrm{max}}$. For the cases $f_{\mathrm{LO}}$=20GHz and 70.5 GHz, the noise parameter $\Gamma$ was taken to be 0.026 corresponding to a critical current of 100$\mu A$ at 58K. For the case $f_{\mathrm{LO}}$=140GHz, it was not possible to obtain a quantitative agreement with $\Gamma=0.026$. The best fit, shown in the figure, was obtained for $\Gamma=0.06$ indicating that the junction is submitted to an extra noise, equivalent to an effective temperature twice larger than the physical one. The conversion efficiency takes a maximum value of 0.3\% at 20GHz and decreases to 0.02\% at 140GHz. Its overall weak absolute value is due to the low impedance of the junction (approximatively 2$\Omega$) compared to the external impedances $Z_u$ and $Z_0$, a problem that could be overcome by modifying the geometry of the junction. In principle, the impedance of the junction could be increased easily up to 20 $\Omega$ by changing both the width and the thickness of the junction, leading to a much higher conversion efficiency. \\
\indent Although the ion-irradiated Josephson junction fabricated for this study has a characteristic frequency well below any reasonable estimate of the \YBCO gap frequency, several developments can be made to optimize the $I_cR_n$ product \cite{sirena, sirenaJAP2007}. In particular, a higher fluence of irradiation combined with an annealing of the sample should lead to a qualitative improvement  \cite{sirenaAPL2007,sirenaJAP2009}.  Let us also mention that junctions made recently by irradiation through larger slits display  $f_c$ values up to 500 GHz  (Supplementary Material figure 1). This is a promising result although their mixing properties have not been measured yet.\\
\indent In conclusion, we have demonstrated the mixing operation of ion-irradiated \YBCO Josephson junctions up to 140 GHz at 58K in good agreement with the three ports RSJ model. In particular, we have clearly shown the transition between two mixing regimes $f_{\mathrm{LO}}<f_c^{\mathrm{max}}$ and $f_{\mathrm{LO}}>f_c^{\mathrm{max}}$. In the short term, characteristic frequencies of order 500 GHz at T$>$40 K are entirely within reach enabling mixing operation up to $\sim$1THz. In addition,  the natural scalability of the ion irradiation technique \cite{cybart} makes it particularly interesting to implement, next to the mixer, an integrated Josephson local oscillator made of a large number of junctions and whose frequency could be tuned by dc biasing \cite{barbara,darula}.\\

\indent The authors thank L. Olanier for technical support. This work was supported by the international cooperation program MINCyT-ECOS A10E05, the  Emergence program Contract of Ville de Paris and by the R\'egion Ile-de-France in the framework of CNano IdF and Sesame program and  by the PEPS NANANA of CNRS. CNano IdF is the nanoscience competence center of Paris Region, supported by CNRS, CEA, MESR and R\'egion Ile-de-France. \\

\newpage

 \begin{figure}[h!]
\includegraphics[width=10cm]{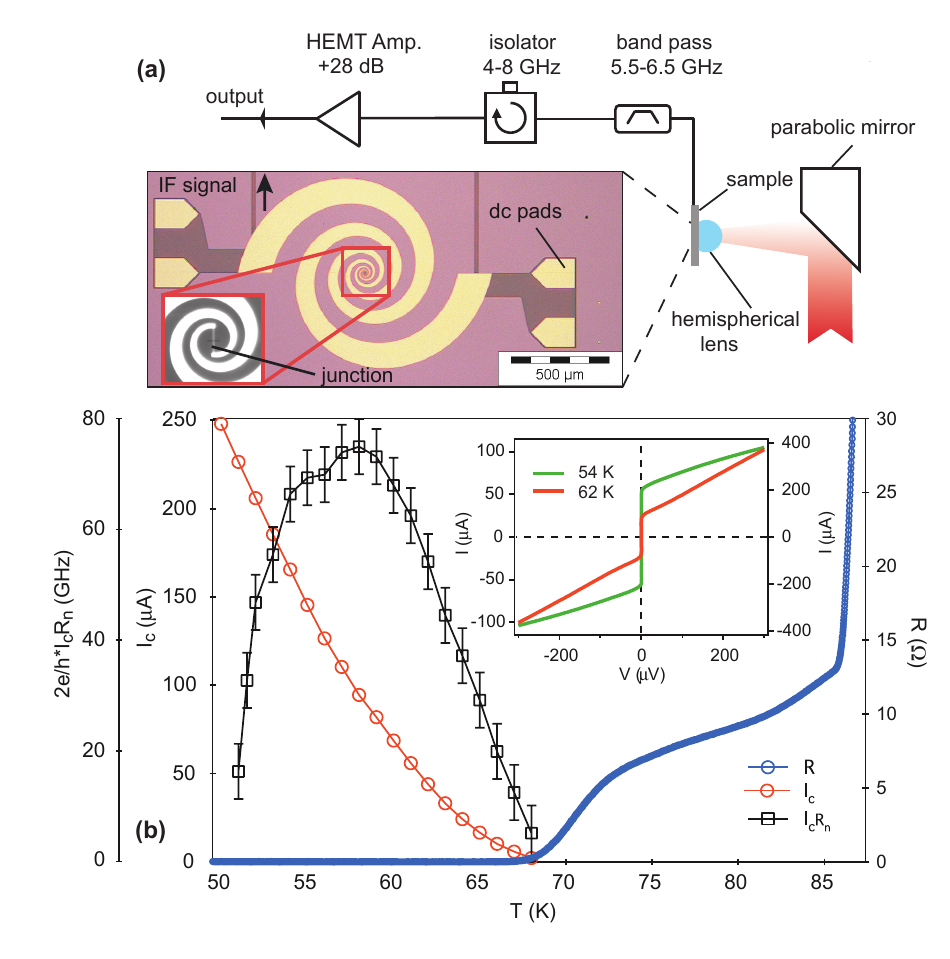}
\caption{ a)  Josephson mixer in its optical and microwave set-up. The junction is embedded in a wide-band spiral antenna (85 GHz - 7 THz).  b) Resistance, critical current and $I_cR_n$ product of the junction as a function of temperature. Inset) I(V) curves at two temperatures below and above $T'_c$ showing the difference between the Josephson regime (upward curvature) and the flux flow regime (downward curvature).}
\end{figure}

\newpage

 \begin{figure}[h!]
\includegraphics[width=10cm]{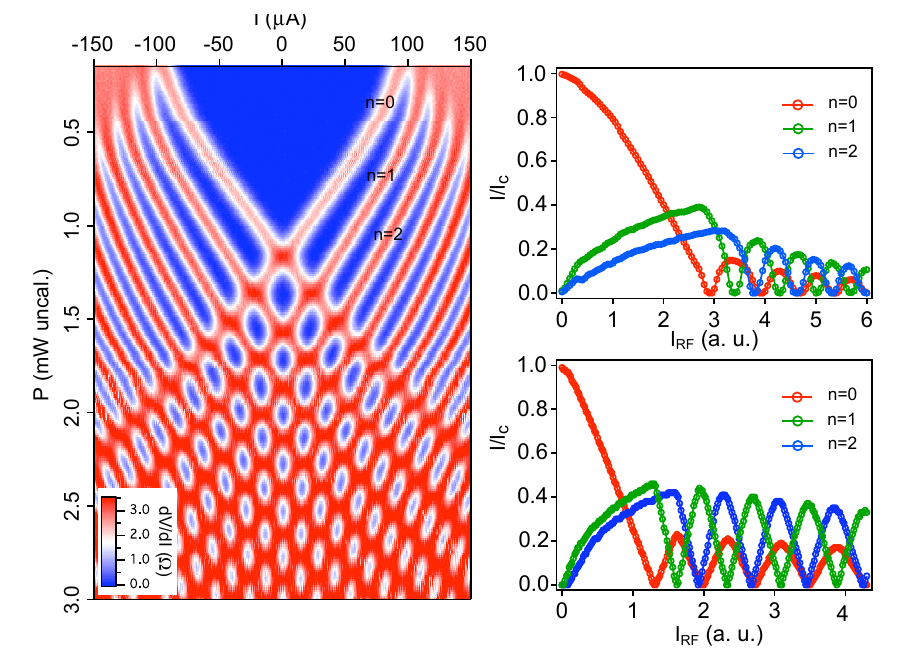}
\caption{ a) Differential resistance of junction (color scale) under a 20 GHz signal as function of bias current and signal power. b) Experimental height of the n=0, n=1 and n=2 Shapiro steps as a function of signal current measured at T=58K. c) Theoretical height of the n=0, n=1 and n=2 Shapiro steps as a function of signal current obtained from the RSJ model.} 
\end{figure}

\newpage

 \begin{figure}[h!]
\includegraphics[width=10cm]{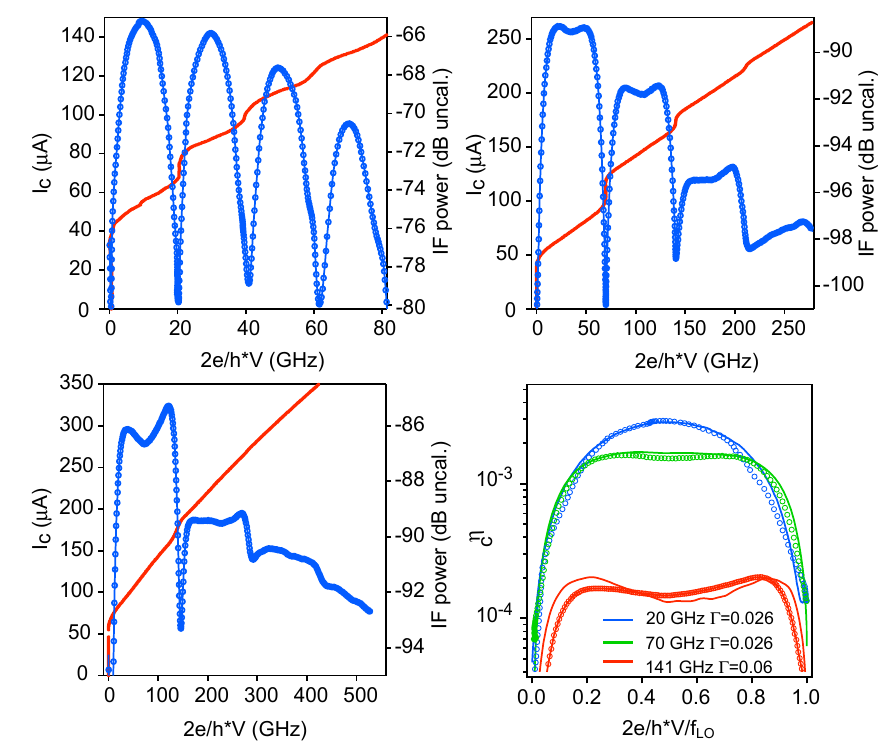}
\caption{: a,b,c) Output power at the intermediate frequency (right scale) and dc current (left scale) as function of voltage for three different frequencies : a)$f_{\mathrm{LO}}$=20GHz and $f_s$=14.5GHz; b) $f_{\mathrm{LO}}$=70.5GHz and $f_s$=76GHz and c) $f_{\mathrm{LO}}=140GHz$ and $f_s$=145.5GHz. The power of the signal has been set to approximatively one thousandth of the power of the local oscillator. d) Comparison between experimental (dots) and theoretical (full lines) conversion efficiency $\eta_C$. The noise parameter $\Gamma$ used for the calculation is indicated on the figure.}
\end{figure}

\newpage

\large{\textbf{Supplementary Material}}\\
\normalsize

\textbf{Irradiation through slits of different widths}\\

 The junction presented in the article is defined in a 2 $\mu$m wide, 70nm thick, \YBCO superconducting channel by irradiating through a 20nm wide slit with 100keV oxygen ions at a fluence of 3$ \times$10$^{13}$ at/cm$^2$.We have also made junction irradiated through 30 nm, 40 nm and 100 nm wide slits. Supplementary figure 1 shows that these junctions exhibit a higher $I_cR_n$ product. In particular, junction irradiated through a 100 nm slit has a maximum characteristic frequency of approximatively 500 GHz, which should enable mixing operation up to 1THz at temperature around 40K. In addition, these junctions have a higher resistance $R_n$ and are therefore easier to match with the antenna and readout circuit impedances.
      \begin{figure}[h!]
\includegraphics[width=10cm]{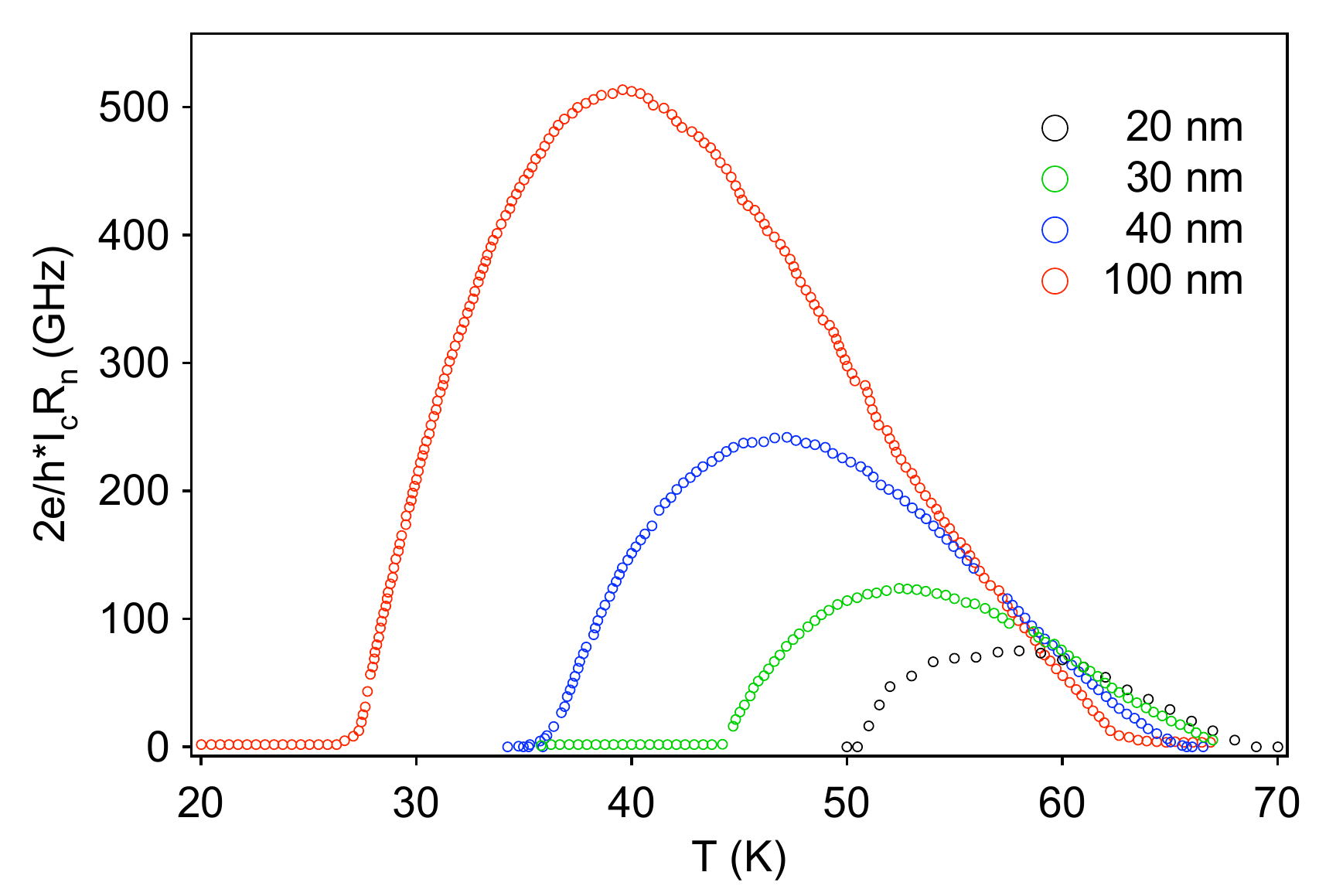}
\caption{$I_cR_n$ product as a function of temperature for junctions irradiated through different widths of the slit.}
\end{figure}

\end{document}